# SPECTRAL CURVE FITTING FOR AUTOMATIC HYPERSPECTRAL DATA ANALYSIS

ADRIAN BROWN[1]

## ABSTRACT

A method for automatic curve fitting of hyperspectral reflectance scenes is presented. Highlights of the method include the ability to detect small relative shifts in absorption band central wavelength, the ability to separate overlapping absorption bands in a stable manner, and relatively low sensitivity to noise. A comparison with techniques already available in the literature is presented using simulated spectra. An application is demonstrated utilizing the Short Wave Infra Red (SWIR – 2.0-2.5 $\mu$m or 5000-4000cm$^{-1}$) region.  A small hyperspectral scene is processed to demonstrate the ability of the method to detect small shifts in absorption wavelength caused by varying white mica chemistry in a natural setting.

---

[1] Australian Centre for Astrobiology, Macquarie University. Email: abrown@els.mq.edu.au, Website: http://aca.mq.edu.au/abrown.htm, now at abrown@seti.org

# INTRODUCTION

Hyperspectral data analysis is often viewed as a statistical pattern recognition problem in a three dimensional hyperspace, often envisaged as a hyperspectral data cube. These methods usually begin with a statistical approach to find image 'end members'. Statistical methods range from simple hierarchical cluster analysis to complex methods such as minimum noise transforms and the RSI 'hourglass' method [1-3].

In contrast to these statistical methods, a more traditional spectroscopic approach can be taken, where the analyst concentrates on recognising absorption band shapes in each individual spectrum. This has the advantage of a direct connection between the reflectance spectrum of the target pixel and its chemical composition. When the hyperspectral dataset is viewed in this way, classifications are made according to underlying physical properties, rather than simply on their similarity to other pixels in the dataset.

Statistical methods have been demonstrated to have a place in the field of hyperspectral analysis; however we present here a tool which automates the processing of hyperspectral data in the manner of traditional spectral analysis.

Hyperspectral curve fitting methods immediately confront the challenge of modeling multiple overlapping absorption bands with relatively low spectral resolution. Typically, a hyperspectral spectrum has 100-250 measurement points (channels), with a spectral resolution per channel of 10-20$\eta$m and similar sampling intervals [4, 5]. Typical modern day laboratory spectroscopic instruments have a spectral resolution of 1.25$\eta$m (2cm$^{-1}$) and similar sampling intervals [6]. The order of magnitude difference in spectral resolution means that hyperspectral spectra are not as definitive as laboratory spectra, and curve fitting techniques from the laboratory cannot simply be transferred to hyperspectral datasets.

Reflectance spectra of solid surfaces can be modeled as a series of deep dips due to absorption bands, superimposed on a background continuum [7]. The central frequency at which absorption bands are detected is of most interest to spectroscopists, since these frequencies are indicative of the mineral types present [8, 9].

Derivative spectroscopy can be used to approximate the locations of absorption bands using second and higher degree differential spectra [10, 11]. Fifth order derivatives have been demonstrated to give the most accurate [12].

Curve fitting of spectra can estimate not only the central wavelength of the absorption band, but also their width and amplitude of the absorption band (Figure 1). This is typically achieved using an iterative least squares step [13].

The absorption band dips can be modeled using Gaussian [14, 15], Lorentz [16], or mixed Gaussian-Lorentz (Voight) curves [17]. These curves are all symmetrical shapes that decay from a central peak (Figure 1). Differing decay rates make each curve type better at replicating different physical processes [18].

Overlapping absorption bands are additive when measured in an absorption experiment. Reflectance spectra can be converted to apparent absorbance by taking the logarithm of the reflectance.

apparent absorbance = log10(reflectance)

Natural logarithm is used in the Physics community; base 10 logarithm has been used in the planetary sciences community [14, 15, 19].

Hyperspectral sensors typically report spectra in units of $\mu$m or $\eta$m (wavelength), however plotting in energy space in $cm^{-1}$ (wavenumbers) has greater physical basis and eliminates asymmetry due to display on a constant interval wavelength abscissa [20].

frequency ($cm^{-1}$) = 10000/ Wavelength($\mu$m)

The mathematics surrounding our curve fitting algorithm will be briefly discussed, followed by analysis of a noise free and noisy synthetic spectrum. Fits using Gaussian, Lorentzian and Voight curves will be compared. Finally, a real world application will be presented.

## METHODOLOGY

If we consider a continuum removed reflectance spectrum [21], recorded at $N$ discrete points, as a one dimensional vector **R**, in wavelength space:

**R**($\lambda$) = R($\lambda_1$), R($\lambda_2$), … R($\lambda_N$)

We may convert to apparent absorbance **A** by taking the base 10 logarithm, as described above, and multiply the spectrum by -1 in order to make the absorption features 'positive':

**A**($\lambda$) = -log10(**R**($\lambda$))

we convert to energy space, where v is frequency, as described above:

**A**(v) = 10000/**A**(λ)

If we consider the absorption bands to be Gaussian in shape, we may model **A**(v) using a series of absorption bands of the following general formula:

$$G(v) = \alpha \cdot \exp\left(-(v - v_0)^2 / 2\sigma^2\right)$$

where α is the amplitude (height) of the Gaussian, $v_0$ is the central frequency, and σ is the full width at half maximum.

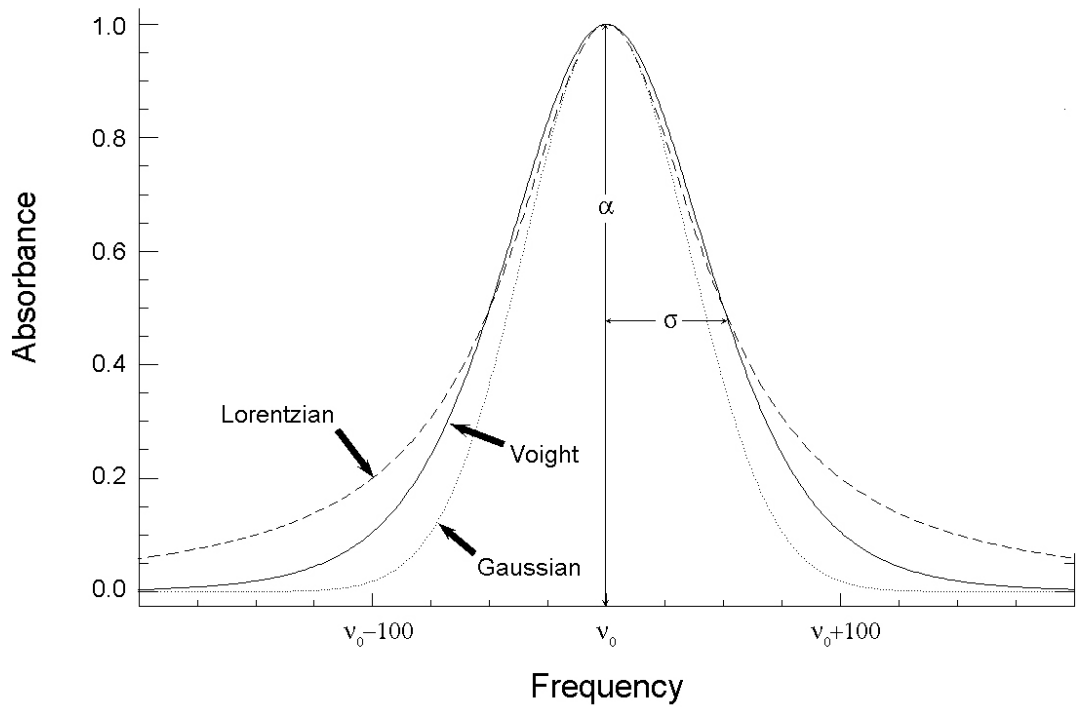

Figure 1 – Comparison of Gaussian, Lorentzian and Voight functions describing an ideal, symmetrical absorption band. For each function, α=1, σ=100. The Voight function has been generated as described in the text, with β=0.5.

If we wish to model the absorption bands as Lorentzian in shape, we may use the formula:

$$L(v) = \alpha \cdot \left(\sigma / ((v - v_0)^2 + \sigma^2)\right)$$

A Lorentzian shape is often applicable for the absorption bands of gases [18], bought about by collision broadening of line spectra. A Gaussian shape is often superimposed by the effects of instrumental smearing (due to channel bandpass functions), and also by Doppler broadening [22]. Thus it is often desirable to model the absorption bands as a multiplicative or additive combination of Gaussian or Lorentzian shapes – this is then a Voight profile. An alternative approach, similar to a Voight profile, was suggested by [23] and is adopted here.

If we model the absorption bands using a function **K**, of the type:

**K**(v) = α . ( 1 / ( 1+ $\beta^2\Psi(v)^2$ )^( $1/\beta^2$ ) )

where $\Psi(v) = (v - v_0) / \sqrt{2}\sigma$

The parameter β is then a reflection of how 'Gaussian-like' or 'Lorentzian-like' our function is. We call it the Voight-likeness parameter. Clearly, if β = 1, then **K** = **L** and we have a Lorentzian shape. If β = 0, then our expression for **K** resembles a Gaussian shape.

This is shown if we use the Taylor's theorem expansion for ln(1+x),

f(x) = ln(1+x) = x − $x^2/2$ + $x^3/3$ − $x^4/4$ + …

Dividing by α, taking the natural logarithm of **K** and expanding, we get:

ln (**K** / α) = − $\beta^{-2}$ ln (1+ $\beta^2 \Psi^2$) = − $\Psi^2$ + 1/2 $\beta^2 \Psi^4$ − 1/3 $\beta^4 \Psi^6$ + 1/4 $\beta^6 \Psi^8$ − …

Now, we let β = 0 and take the exponential of both sides to get

**K** = α . exp (−$\Psi^2$)

And hence, **K** = **G**.

Using our approximation of an absorption band, **K**($v_0$, α, σ, β), we may now approach the problem of fitting our curves to the absorption spectra **A** in the following way.

Hyperspectral datasets usually do not have enough measured points to carry out least squares analysis. The number of measured points must exceed the number of free parameters [13]. For example, with 4 free parameters per absorption band, only five bands can be modeled with 21 measured points. Thus it becomes necessary to interpolate to achieve a greater number of measurement points. We adopted an interpolation procedure similar to that described by Dyn et al. to model computer graphics curves. This interpolation approximates a cubic spline fit [24]. This interpolation procedure decreases the signal to noise (SNR) of the spectrum, so the number of interpolated points should be kept to a minimum. We found two interpolation runs sufficient to achieve good least squares fit, but this is clearly application dependent. Laboratory spectra should require no interpolation, for example.

For each absorption band, we must find four parameters, $v_0$, $α$, $σ$ and $β$. The location of each peak, $v_0$, is found by using an adaptive width sliding window Savitzky-Golay routine to model a sixth degree polynomial, and calculate the fifth, fourth and second derivatives of this polynomial [25]. A peak is located where the following conditions are fulfilled [12] :

$A^V(v)=0$, $A^{IV}(v)>0$ and $A^{II}(v)<0$

The superscript in Roman numerals indicates the order of the derivative of the absorption spectra.

The width of the sliding window is crucial to the accuracy of this peak finding method. If the bands are wide, the sliding window should be large to accurately fit the Savitzky-Golay polynomials. The opposite is the case for narrow bands. Huguenin's method of choosing this width is proprietary. Our method uses the width of second derivative positive and negative going peaks to approximate the width for each point in the spectrum.

Once the $M$ central peaks are located, we can approximate an initial solution to the other three parameters ($α$, $σ$ and $β$) for each band, **K**.

To estimate $α$, we use linear interpolation to estimate the value of **A**(v) at $v_0$ for each of the $M$ peaks:

$α_m = \mathbf{A}(v_{0m})$    where m = 1, 2, …, M

To estimate the full width at half maximum, $σ$, we look either side of the peak value, for the closest point in the absorption spectrum **A**(v) where the value has fallen to half of the value $α$. The distance either side is calculated, and the smallest distance is taken as our approximation to $σ$.

The value of $β$ for each absorption band is initially set to a value of 0.5, to approximate an equally mixed Lorentzian-Gaussian curve [26].

Once these initial estimates of the parameters are obtained, we use a two stage iterative least squares method to refine them.

The iterative least squares method minimizes the following relationship:

Min $\|\mathbf{A} - \mathbf{K}_{total}\|$

where $\mathbf{K}_{total} = K_1 + K_2 + \ldots K_M$

There are many such methods available [13, 27-29] - we adopted the Levenberg-Marquardt scheme [13]. After the initial estimate and first least squares refinement, the residual error, $\chi^2$ (chi-squared) is calculated.

$$\chi^2 = \left( \frac{1}{N} \left( \sum_{n=1}^{N} [A(v_n) - K_{total}(v_n)]^2 \right) \right)^{1/2}$$

If the residual is below a certain threshold, iteration is ended. Threshold values for $\chi^2$ are largely dependent on the dataset being examined, and we recommend experimentation with this value. Values of 0.01 have been quoted in the spectroscopy literature [26] and were found to give acceptable results, however we found values of $1 \times 10^{-3}$ gave optimal results for our HyMap dataset, which is discussed below. If the $\chi^2$ value does not dip below the threshold after 20 iterations, the loop is exited.

In the first stage of iterative refinement, the $v_0$ and $\beta$ values are pinned at their initial values, and $\alpha$ and $\sigma$ are allowed to iterate to a close solution. During the second stage, all parameters are free to vary.

This procedure utilizes the best of two techniques that are commonly used separately - the high order derivative techniques, and a staged iterative least squares curve fitting technique. The two stage enhancement affords stability, since the initial guesses for $\alpha$ and $\sigma$ for overlapping bands are often variable in accuracy. It also enhances the speed of the curve fitting overall since the starting estimates for the second stage require little adjustment.

The curve fitting algorithm was developed using IDL 6.0 from RSI (http://www.rsi.com).

## SYNTHETIC SPECTRA MODELING

A continuum removed reflectance spectrum was generated synthetically, and then analysed to see if the parameters could be retrieved using our curve fitting algorithm.

The parameters for the synthetic spectrum were chosen to be the same as those used by Huguenin to produce his results [12]. Since Huguenin's spectra were already absorption spectra in energy space, conversion from reflectance was not required. The curve parameters appear in Table 1. These values were chosen as three pairs of overlapping Gaussian shaped bands, very close to the limit of resolution from each other. Huguenin calculated the limit of resolution of two overlapping curves to be

$v_2 - v_1 > 0.56W$

where W is the widest band's full width at half maximum, $v_2$ and $v_1$ are their respective central frequencies.

| Band | $v_0$ Wavelength position, cm$^{-1}$ | $\sigma$ Half Width, cm$^{-1}$ | $\alpha$ Strength, %A |
|---|---|---|---|
| 1 | 9500 | 2355 | 30 |
| 2 | 11500 | 3040 | 42 |
| 3 | 14500 | 1990 | 30 |
| 4 | 16000 | 2150 | 34 |
| 5 | 18500 | 2033 | 60 |
| 6 | 20500 | 2150 | 80 |

Table 1 – Parameters for Huguenin's synthetic Gaussian absorption bands

The sum of errors, $\Sigma_{error}$ was calculated by summing the absolute difference between the final result for $v_0$ of each absorption band, and the known, synthetic $v_0$.

$$\Sigma_{error} = {}^M\Sigma_{m=1} \parallel v_0 \text{ (calculated)} - v_0 \text{ (synthetic)} \parallel$$

Huguenin did not specify how many measurement points he used, therefore we used a range of values for *N*. Since parts of Huguenin's technique is proprietary, we cannot be sure how closely we approximated his process. He quoted a value for $\Sigma_{error}$ of 143 (his Table 3) for this scenario.

We tested our our method with our standard initial value of β=0.5, and also at β=0.1, a value closer to the synthetic Gaussian shapes. We expected that starting close to the expected values would result in smaller final errors. Sampling resolutions were calculated assuming a measurement range of 15000cm$^{-1}$ (from 5000-20000 cm$^{-1}$). The results are shown at Table 2.

| | | $\Sigma_{error}$ | | |
|---|---|---|---|---|
| N | Sampling resolution, cm$^{-1}$ | Huguenin's Fifth derivative method | This paper, initial β=0.1 | This paper, initial β=0.5 |
| 88 | 170.5 | 148 | 61 | 39 |
| 100 | 150 | 153 | 48 | 40 |
| 500 | 30 | 156 | 53 | 24 |
| 1000 | 15 | 155 | 79 | 41 |

Table 2 – Comparison of results for derivative methods and curve fitting algorithms. $\Sigma_{error}$ is the sum (over 6 absorption bands) of the absolute errors in calculated values for $v_0$.

The $\Sigma_{error}$ figures for different numbers of measurement points illustrates several points.

- The addition of a least squares refinement stage has the capacity to significantly improve (up to 3 times in this scenario) upon the pure derivative technique.
- Although the synthetic curves were Gaussian, better peak position fits are achievable using an initial Voight-likeness parameter half way between Gaussian and Lorentzian behavior, compared to fits where the Voight likeness parameter was started off closer to a Gaussian shape. This preference for starting values of β =0.5 was not reflected in final values of the β parameter – all of which settled below 0.5 and all (bar two) below 0.01 after iterative refinement.
- Little is gained by an order of magnitude increase in *N* from 100 to 1000, but a value of *N*=88 is required to reach this stable plateau, since below this value, an extra peak is found in this scenario of 6 closely spaced bands. For some values of *N* below 88 the correct number of bands is sometimes found, but in the wrong positions – generally giving $\Sigma_{error}$ values around 700-1000.

In order to test the effectiveness of our interpolation routines, we used Huguenin's scenario with two interpolation runs, and compared the $\Sigma_{error}$ to our previous results. These results appear in Table 3.

| N | Sampling resolution, $cm^{-1}$ | Points after interpolation | $\Sigma_{error}$ | | |
|---|---|---|---|---|---|
| | | | Huguenin's Fifth derivative method | This paper, initial β = 0.1 | This paper, initial β =0.5 |
| *1 interpolation run* | | | | | |
| 50 | 300 | 99 | 146 | 86 | 43 |
| *2 interpolation runs* | | | | | |
| 33 | 454.5 | 129 | 153 | 103 | 56 |
| *3 interpolation runs* | | | | | |
| 33 | 454.5 | 257 | 151 | 105 | 65 |
| *4 interpolation runs* | | | | | |
| 33 | 454.5 | 513 | 151 | 61 | 69 |
| *5 interpolation runs* | | | | | |
| 33 | 454.5 | 1025 | 149.6 | 48.51 | 150 |

Table 3 – Comparison of results for derivative methods and curve fitting algorithms using two interpolation runs. $\Sigma_{error}$ is the sum (over 6 absorption bands) of the absolute errors in calculated values for $V_0$.

For values of N less than those shown in Table 3, extra peaks were 'found' by our curve fitting method. Although good fits were still achieved, an extra peak means the errors are not simply assessable. We regard this as a failure under this scenario, although it would still give useful information as long as the 'found' peaks were small and did not overlap much with real peaks.

The results of running the interpolation routines justifies their use – values for N as low as 33 still gave excellent reproductions of peak positions in this challenging scenario.

The trend for better results from mid range values for β is reversed as more interpolation is used. This suggests that our interpolation is introducing even greater Gaussian tendencies into our data.

It was found that increasing the number of interpolations past 2 did not give any better performance in peak finding for this scenario. It did not escape our notice that the number of final points after interpolation (129 points) is similar, but slightly larger than the number of points required for good reproduction of peak positions without interpolation (88 points). This is intuitively reasonable, though currently we do not have an analytical understanding of the relationship between these two factors.

Since more interpolated points increases computation time without appearing to improve performance (although this has not been exhaustively tested under other scenarios), we use a standard of two interpolation runs for the remainder of this paper.

*Noise Sensitivity testing*

To test the noise sensitivity of our method we generated another synthetic spectrum simulating the SWIR (4000-5000cm$^{-1}$) spectrum of chlorite obtained by a hyperspectral instrument. The parameters of the 4 bands are given in Table 4.

| Band | $v_0$ Wavelength position, cm$^{-1}$ | σ Half Width, cm$^{-1}$ | α Strength, %A | β Voight likeness |
|---|---|---|---|---|
| 1 | 4550 | 25 | 1 | 0.1 |
| 2 | 4410 | 60 | 1.5 | 0.1 |
| 3 | 4315 | 50 | 2 | 0.1 |
| 4 | 4190 | 50 | 2.5 | 0.1 |

Table 4 – Absorption band parameters for synthetic chlorite spectrum

The synthetic spectra were then modeled by our curve fitting algorithm with varying amounts of noise, and varying numbers of measurement points. RMS noise was calculated in reflectance space. Two interpolation runs were used for each scenario. The Σ$_{error}$ was calculated by summing the difference between the calculated and synthetic central wavelengths, $v_0$. The results appear at Table 5. Because the central frequencies are smaller here than our first example, the results are not directly comparable with Tables 2 and 3.

|  | RMS noise (in reflectance space) | | |
| --- | --- | --- | --- |
| *Real measurement points, N* | $2 \times 10^{-5}$ | $10^{-5}$ | $10^{-6}$ |
| 20 | 9 | 11 | 3 |
| 45 | - | 3 | 1 |
| 89 | - | - | 4 |

Table 5 – Synthetic spectral modeling results for varying amounts of noise and numbers of measurement points. Values given are for $\Sigma_{error}$, calculated as described in the text. Two interpolation runs were carried out for each scenario. Note that since the peaks were of smaller frequency, the error values are not directly comparable with Tables 2 and 3. A dash means that extra peaks were found so we cannot asses the error.

The noise characteristics of our curve fitting algorithm may be understood in the following way. As explained earlier, the appearance of an extra band is regarded as a failure. When we add more measurement points, we add more noise to each point and more peaks are apparent to the curve-fitting algorithm.

Flat response regions of a spectrum with added noise take on the appearance of small peaks. Since hyperspectral data often show peaks with just one measured point, Fourier smoothing denoising techniques [12] cannot be applied without the risk of eliminating small absorption bands.

The appearance of spurious peaks does not frustrate spectral analysis – these can often be eliminated by careful amplitude thresholding, except where they appear in overlap with other bands. Our analysis here highlights the critical nature of the noise characteristics of hyperspectral sensors.

Figure 2 displays an example of the synthetically generated chlorite absorption spectra with RMS noise of $10^{-5}$, along with the pure derivative fit, and fit after least squares refinement.

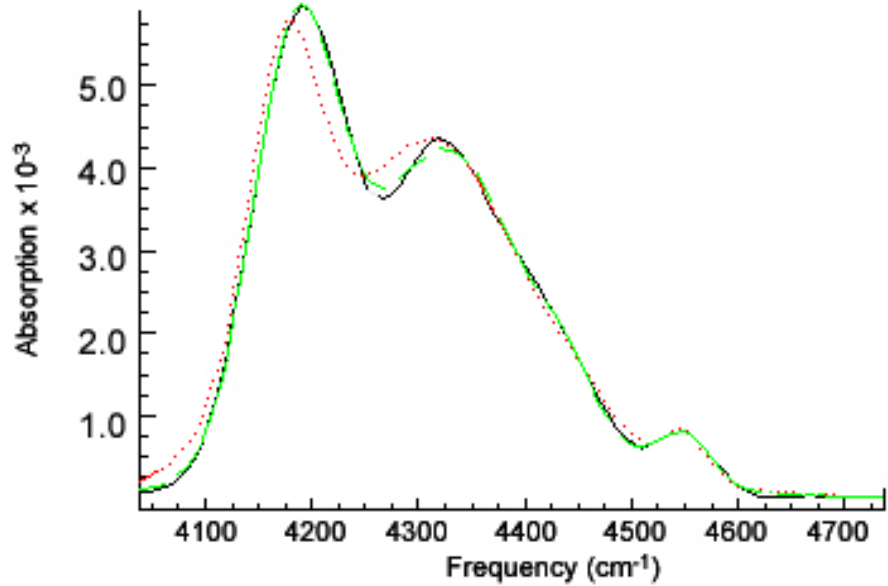

Figure 2 – Black line shows synthetic chlorite absorption spectrum with RMS noise = $10^{-5}$. Red dots indicate the fit before iterative refinement. Green curve shows final fit.

In order to test the difference between Gaussian, Lorentzian and Voight curve fitting, we fitted a real chlorite SWIR spectrum (obtained from a HyMap dataset) with each type of curve. No noise was added, and two interpolation runs were carried out on an initial number of measurements $N=23$. Table 6 shows the $\chi^2$ values for each type of curve after our modeling technique.

|  | Curve Type | | |
| --- | --- | --- | --- |
|  | Gaussian | Lorentzian | Voight |
| $\chi^2$ | 0.018 | 0.026 | 0.009 |

Table 6 – $\chi^2$ values for a real chlorite spectrum, modeled using exclusively Gaussian, exclusively Lorentzian or Voight type curves.

These results shown in Table 6 were typical of SWIR spectra from our dataset, which is discussed below. The Lorentzian only curve fits performed the poorest, and the more flexible Voight curves provided the best overall fit. The Lorentzian shapes were probably not assisted by our choice of continuum removed spectra – the slow fade out to the edges of Lorentzian shapes does not match the sharp edges of straight line continuum removed spectra.

# APPLICATION TO HYMAP DATASET

We tested our curve fitting algorithm on a small subset of a HyMap dataset [5]. The dataset was collected in fine conditions in October 2002 over the Pilbara region of Western Australia [30]. The region is an Achaean granite greenstone terrain, displaying greenschist facies indicator minerals such as chlorite, hornblende, albite, and actinolite. Hydrothermal activity has emplaced muscovite rich veins and horizons throughout the terrain. The small scene we examined is believed to represent a volcanic edifice, rich in muscovite of varying Al/Si ratio, and surrounded by a chlorite-hornblende rich greenstone region [31]. Pixel resolution is approximately 5m on the ground.

We limited our investigation to the Short Wave Infrared (SWIR) region. This region is particularly compelling due to its position on a high reflectance plateau. The SWIR region is largely unobscured by atmospheric effects. Band saturation is uncommon in the SWIR region due to generally high reflectance [32]. The absorption bands in the SWIR region provide diagnostic information about the presence of hydrous phyllosilicates (such as micas, chlorites, serpentine, talc, etc.) which can be used to map hydrothermally altered zones [33]. Finally, curve fitting the SWIR is attractive due to relatively high signal to noise (SNR) achievable in this region with modern instruments (for example, HyMap SNR varies from 250-1000 in the SWIR [5]).

Before the analysis, atmospheric correction was carried out using ATREM [34]. The background continuum was removed using automatic continuum removal procedures [21] in order to leave only residual absorption band shapes suitable for modeling. For this study, straight line approximations were made for continuum removal.

Since we are using the SWIR region of the HyMap dataset, we have 23 available channels, or measurement points, from 2.0932μm to 2.4768μm. Two interpolating steps were carried out.

Using the curve fitting algorithm, maps were generated of the Al-OH combination (bend δ + stretch ν) absorption band near 2.19μm. This absorption band is indicative of the presence of white mica, such as muscovite. A similar band is also attributable to kaolinite, but this always occurs with another band at 2.16 μm, and in order to find just white mica, we eliminated any spectra bearing 2.16 and 2.19μm bands.

The results of mapping the 2.19 μm absorption band are shown in Figure 3. Maps are generated of all four parameters, $v_0$, α, σ and β.

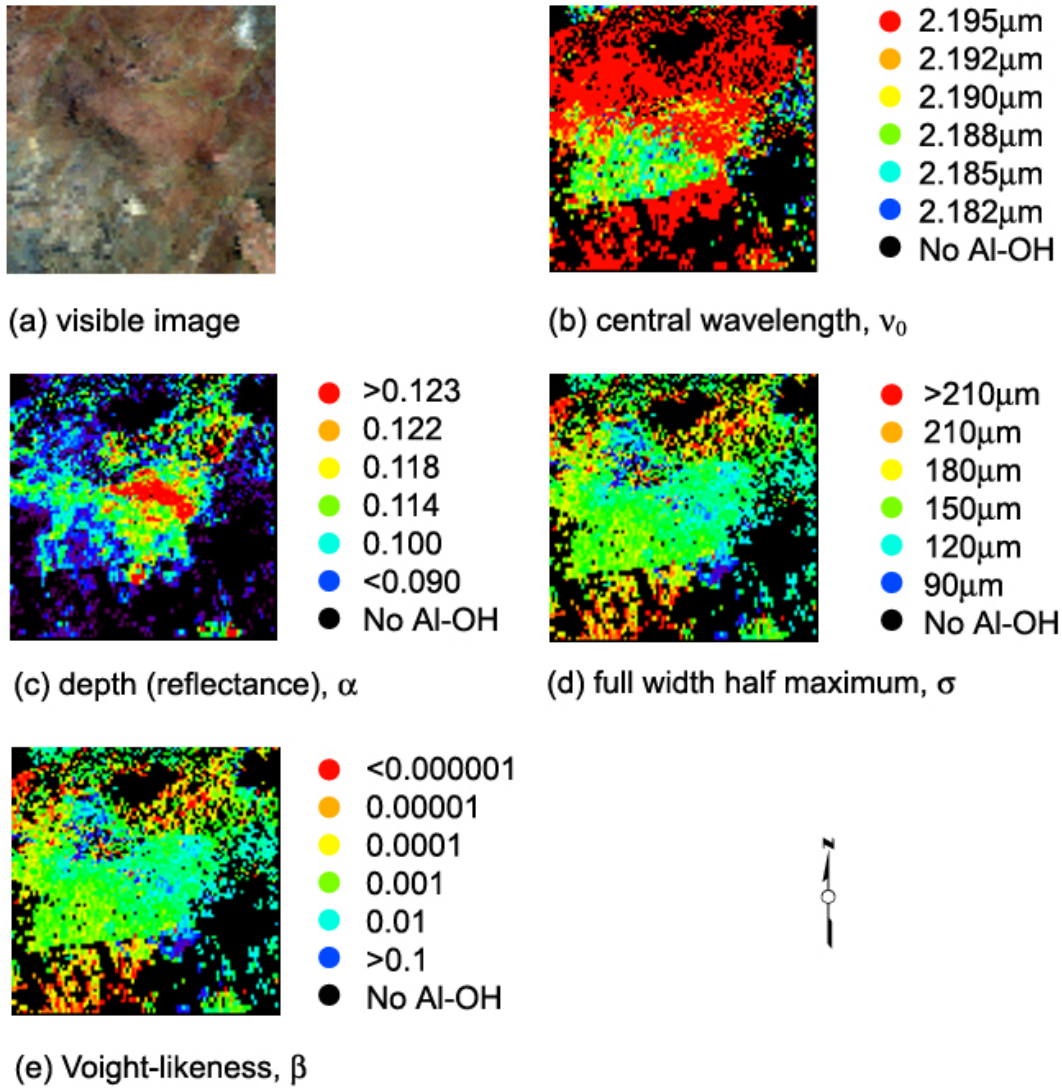

Figure 3 – Maps of Al-OH band parameters. (a) visible image (b) central wavelength, $v_0$ (c) depth of absorption band, α (d) full width half maximum, σ (e) gaussian-lorentzian figure, β. Values given are the bottom of each range defined by different colors.

The map of central wavelengths (Fig 3b) clearly shows short wavelength Al-OH in a swathe in the middle of the image. This is attributable to high temperature Tschermak substitution [35] in white mica within the volcanic vent. The map of amplitude (Fig 3c) shows a small band of high amplitude Al-OH in the middle of the image, but not coincident with the regions of low wavelength. This is broadly attributable to the abundance of mica in these locations [6]. The FWHM map (Fig 3d) shows the large preponderance of FWHM at around 150µm, although this is variable on the fringes of the white mica regions. The Voight-likeness parameter shows a fairly uniform band around $\beta=0.001$.

# DISCUSSION

<u>Voight, Lorentzian or Gaussian?</u> Our research so far has not demonstrated conclusively which type of curve is most suitable for modeling absorption bands in the SWIR region. Our method shows Voight type curves are more flexible and give a better $\chi^2$ fit. The physical significance of the Voight-likeness parameter $\beta$ has not yet been fully evaluated, but the ability to explore this parameter in future research makes this a valuable addition to our curve fitting algorithm.

<u>Asymmetry as a curve parameter</u>. In addition to the parameters $\alpha$, $v_0$, $\sigma$ and $\beta$, the asymmetry of an absorption band can be modeled; however asymmetry can also be caused by hidden overlapping bands. In this study all absorption bands in the SWIR region are assumed to be symmetric and asymmetry is modeled by the fitting of extra absorption bands in the wings of major absorption bands.

<u>Continuum removal or continuum modeling?</u> In order to keep a cap on the free parameters, we chose to use continuum removed spectra, rather than modeling the continuum. This tends to favor a Gaussian shape rather than Lorentzian shapes (since Gaussian curves fall off more sharply, see Figure 1). Future work will examine the benefits of continuum modeling or non iterative modeling by cubic or higher order polynomials.

<u>Potential overfitting of data</u>. The essence of using an iterative fitting algorithm to model reflectance spectra is the choice of the degrees of freedom employed in the fit. A unique solution to the least squares fitting is not possible [36]. A larger number of absorption bands, and parameters (and hence degrees of freedom), will result in a better fit. Whether the addition of extra absorption bands is justified physically is always open to question. The method outlined in this work, employing a fifth derivative peak fitting algorithm to fix the number of peaks, avoids 'overfitting' of the data with extra absorption bands where they may not be justified (eg. remote hyperspectral data).

<u>Trade offs for using least squares minimization analysis</u>. It is clear that our method can achieve greater accuracy than derivative techniques. This is mostly attributable to the least squares minimization process. This step is time consuming, and pure derivative methods still have a use as a quick approximation to absorption band central wavelengths.

<u>Application to other regions of the spectrum</u>. Our method is here employed in a limited region of the EM spectrum, namely 2.0-2.5µm. There is nothing inherently limiting the procedure to this region. Our method is also applicable to laboratory or field spectra, making comparison of results with airborne or satellite spectra convenient.

<u>Future directions in hyperspectral sensors</u>. Our application has highlighted the advantages of higher spectral sampling resolution for future hyperspectral sensors. More measurement points would obviate the need for interpolation runs. Our analysis method would immediately benefit from a higher number of spectral channels, even if the channel bandpasses were not improved (although this, too, is of course desirable!). Our analysis has demonstrated the benefits of improvements in hyperspectral instrument signal to noise ratios.

<u>Expert system analysis</u>. Our method returns the number of absorption bands, *M*, and for each band, the width σ, amplitude α, central wavelength $v_0$ and Voight-likeness parameter β. This data can be used as input to an expert system for automatic mineral analysis. Such a system has been described by Clark et al [37].

## CONCLUSION

We have presented a technique for automatic curve fitting of hyperspectral reflectance datasets and provided an application. Our method works by combining fifth derivative peak finding with a multiple stage iterative least squares refinement.

The method provides a repeatable, stable modeling method, preserving the complete shape of the spectra. It employs a Voight-type absorption band curve model. It provides the capability to explore maps of band parameters width σ, amplitude α, central wavelength $v_0$ and Voight-likeness parameter β – all of which can be used as an input to an automatic mineral recognition expert system.

## ACKNOWLEDGEMENTS

The generous provision of the HyMap dataset by HyVista and CSIRO is gratefully acknowledged.